# Title
Transcription factor target site search and gene regulation in a background of unspecific binding sites

# Authors
J. Hettich and J. C. M. Gebhardt*

Institute of Biophysics, Ulm University, Albert-Einstein-Allee 11, 89081 Ulm
*To whom correspondence should be addressed: christof.gebhardt@uni-ulm.de

# Abstract
Response time and transcription level are vital parameters of gene regulation. They depend on how fast transcription factors (TFs) find and how efficient they occupy their specific target sites. It is well known that target site search is accelerated by TF binding to and sliding along unspecific DNA and that unspecific associations alter the occupation frequency of a gene. However, whether target site search time and occupation frequency can be optimized simultaneously is mostly unclear. We developed a transparent and intuitively accessible state-based formalism to calculate search times to target sites on and occupation frequencies of promoters of arbitrary state structure. Our formalism is based on dissociation rate constants experimentally accessible in live cell experiments. To demonstrate our approach, we consider promoters activated by a single TF, by two coactivators or in the presence of a competitive inhibitor. We find that target site search time and promoter occupancy differentially vary with the unspecific dissociation rate constant. Both parameters can be harmonized by adjusting the specific dissociation rate constant of the TF. However, while measured DNA residence times of various eukaryotic TFs correspond to a fast search time, the occupation frequencies of target sites are generally low. Cells might tolerate low target site occupancies as they enable timely gene regulation in response to a changing environment.

## Introduction

Gene regulation is mediated to large extend by transcription factors (TFs) interacting with specific target sites on DNA (van Hijum et al., 2009; Venters and Pugh, 2009). Frequently, TFs perform their regulatory function as homo- or heterodimers (Amoutzias et al., 2008), for example nuclear receptors (Horwitz et al., 1996). Once bound, TFs recruit activating or repressing cofactors, chromatin remodelers and the transcription machinery (Coulon et al., 2013). This recruiting activity and as a consequence the regulatory function of TFs is more efficient the larger their probability to occupy the specific target site (occupation frequency) is (Bain et al., 2012; Clauß et al., 2017).

The time a TF needs to find its specific target site in part determines the response time of a gene. It has been proposed that the TF target site search is accelerated if the TF not only tests individual sites on DNA by freely diffusing through the nucleoplasm (3D search), but in addition slides along DNA in an unspecific binding mode (1D search)(Berg et al., 1981; Winter et al., 1981). Indeed, binding of proteins to and sliding along unspecific DNA sequences is a common observation in reconstituted systems (Blainey et al., 2006; Gorman et al., 2010; Kabata et al., 1993; Kim and Larson, 2007) and has been observed in prokaryotes (Elf et al., 2007; Hammar et al., 2012). In mammalian cells, dissociation rate constants from unspecific and specific binding sites are readily accessible by live cell experiments such as single molecule imaging, fluorescence cross correlation or fluorescence recovery after photobleaching (Gebhardt et al., 2013; Mazza et al., 2012), but direct observation of sliding has not yet been possible.

How various parameters such as the ratio of 1D to 3D search mechanisms, crowding, roadblocks or DNA conformation influence the search time has been investigated comprehensively (Bauer and Metzler, 2012; de la Rosa et al., 2010; Gerland et al., 2002; Hu et al., 2006; Klenin et al., 2006; Koslover et al., 2017; Krepel and Levy, 2016; Li et al., 2009; Shvets and Kolomeisky, 2016; Slutsky and Mirny, 2004; Winter et al., 1981; Zabet and Adryan, 2013; Zabet et al., 2013). It is also known that unspecific binding alters the occupation frequency of a gene (Gerland et al., 2002; Hippel et al., 1974; Schoech and Zabet, 2014; Slutsky and Mirny, 2004; Vonhippel and Berg, 1986). Yet it is unclear how both parameters compare under variation of TF-DNA dissociation rate constants. How do TFs distribute over time between their specific target site and the myriad of unspecific sites on DNA (Figure 1)? May a TF find its target site quickly while at the same time occupying it efficiently to optimize gene regulation, or are both requirements contradictory?

To answer these questions, we developed a simplified but powerful state-based formalism including gene promoters of arbitrary state structure. TF target site search times are obtained accurately by solving the Kolmogorov equations for the time evolution to the steady state, and occupation frequencies by considering the steady state distribution. Our formalism is based on dissociation rate constants and thus readily enables calculation of target site search time and occupation frequency using experimentally accessible values. We applied our approach to a simple promoter activated by a single TF and validated our results by comparison with a microscopic search model (Mirny et al., 2009; Shvets and Kolomeisky, 2016). Next, we expanded the activation mechanism by adding a cofactor dimerizing on DNA or a competitor inhibiting specific binding of the TF. We found that both fast target site search and high occupation frequency can be harmonized by adjusting the binding time to the specific target site. Notably, experimentally obtained unspecific and specific dissociation rate constants of various TFs indicate that many TFs indeed are optimized for fast target site search, but not for high occupation frequencies of this site. Activating or competing cofactors allow modulating the optimal conditions, enabling the cell to fine tune the response times and activities of genes in genetic networks.

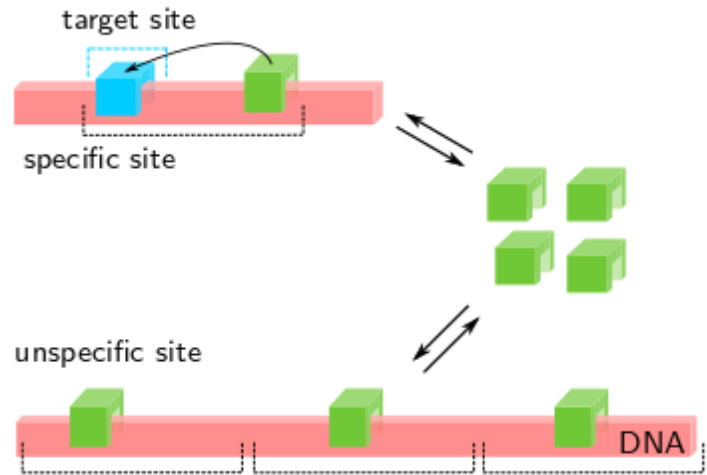

Figure 1: General scheme of a TF binding to DNA. A specific binding site (e.g. a promoter) of arbitrary state structure is embedded in a pool of unspecific binding sites. Drawn is a simple promoter consisting of the target site and surrounding unspecific DNA sequences. Depending on the respective binding times, the TF distributes between the nucleoplasm, unspecific sites and the promoter. If unspecific binding is favored, both the search time and promoter occupancy decrease.

## Materials and Methods

### 1. Mean number of free TFs

*Single TF*
The unspecific binding of the TF crowd is modelled by a chemical reaction scheme in which the rates depend on the number of bound TFs and occupied binding sites (McGhee and Hippel, 1974). Let the total number of unspecific binding sites be *B*. We assume that one binding site BS can only contain one TF at a time. The chemical reaction for binding to unspecific sites is

$$TF + BS \leftrightarrow TFBS \qquad\qquad 1$$

If *l* TFs are bound to DNA, each TF can dissociate independently with the dissociation rate constant $\mu_1$. The association rate constant of TFs is given by the product of free TFs *N-l*, the free binding sites *B-l* and the diffusion limited arrival rate $\lambda_{10}$. The stationary Kolmogorov equation for the probability $p_l$ to find *l* bound TF is

$$0 = \mu_1\left((l+1)p_{l+1} - p_l\right) - \lambda_{10}\left((N-l)(B-l)p_l - (N-l+1)(B-l+1)p_{l-1}\right)$$
$$\text{where } p_l = 0 \text{ for } l < 0 \cup l > M = \min(N,B) \qquad 2$$

This is a recurrence relation which can be solved for $p_l$. This is done by calculating the initial value

$$p_0^{-1} = \sum_{l=0}^{M} \binom{N}{l}\binom{B}{l} l! K_0^l \quad \text{where } K_0 = \frac{l_{10}}{m_1} \qquad 3$$

The expectation value $\bar{n}$ of unspecifically bound TFs is obtained by the derivative

$$\bar{n} = -K_0 \frac{d}{dK_0}\ln(p_0) \qquad 4$$

Since this equation cannot be discussed analytically we next provide an approximation, which aims at revealing the dependencies of the number of unspecifically bound TFs on the arrival and dissociation rates as well as the number of unspecific binding sites. In order to find this approximation, we take the average over Eq. 2 by multiplying with $l$ and performing a summation with respect to $l$ and obtain

$$\frac{\mu_1}{\lambda_{10}}\bar{n} = BN - (B+N)\bar{n} + \overline{n^2} \qquad 5$$

For B, $\mu_1/\lambda_{10} \gg N$ the approximation

$$\bar{n}(\mu_1) = \frac{N}{1 + \dfrac{\mu_1}{\lambda_{10}B}} \qquad 6$$

is obtained. In this approximation the TF is diluted such that no saturation effects can occur. With this approximation we obtain the free number of TFs

$$n_f(\mu_1) = N - \bar{n} = \frac{N}{1 + \dfrac{\lambda_{10}B}{\mu_1}} \qquad 7$$

## TF and competitor

For a second species C the binding to unspecific sites is dependent on the unspecific binding of the TF. We again assume that a unspecific binding site can only be occupied by one molecule. The binding can be written in terms of the chemical reaction

$$TF + BS \leftrightarrow TFBS$$
$$C + BS \leftrightarrow CBS \qquad 8$$

The stationary Kolmogorov equation is

$$0 = \mu_1((l_1+1)p_{l_1+1,l_2} - p_{l_1,l_2})$$
$$-\lambda_{10}((N_1-l_1)(B-l_1-l_2)p_{l_1,l_2} - (N_1-l_1+1)(B-l_1-l_2+1)p_{l_1-1,l_2})$$
$$-\zeta_1((l_2+1)p_{l_1,l_2+1} - p_{l_1,l_2})$$
$$-\Lambda_{10}((N_2-l_2)(B-l_1-l_2)p_{l_1,l_2} - (N_2-l_2+1)(B-l_1-l_2+1)p_{l_1,l_2-1})$$

where $p_{l_1 l_2} = 0$ for $l_1, l_2 < 0 \cup l_1 > M_1 = \min(N_1, B), l_2 > M_2 = \min(N_2, B)$

Here, association and dissociation of the TF are characterized by $\lambda_{10}$ and $\mu_1$, and of the competitor by $\Lambda_{10}$ and $\zeta_1$, and the indices 1,2 with N and l denote TF and C, respectively. This recurrence relation can be solved for p by calculating the initial value

$$p_0^{-1} = \sum_{l_1,l_2=0}^{M_1,M_2} \binom{N_1}{l_1}\binom{N_2}{l_2}\binom{B}{l_1+l_2}(l_1+l_2)! K^{l_1} H^{l_2}$$

where $K = \lambda_{10}/\mu_1$ and $H = \Lambda_{10}/\zeta_1$. The expectation values $\bar{n}_1, \bar{n}_2$ of unspecifically bound TF and competitor are obtained by the derivatives

$$\bar{n}_1 = -K \frac{d}{dK} \ln(p_0)$$
$$\bar{n}_2 = -H \frac{d}{dH} \ln(p_0)$$

Similar to the case of a single TF we provide an analytical solution for the regime in which TFs are diluted. We obtain two equations be averaging over $l_1$ and $l_2$

$$\frac{\mu_1}{\lambda_{10}} \bar{n}_1 = BN_1 - B\bar{n}_1 - N_1(\bar{n}_1+\bar{n}_2) + \overline{n_1(n_1+n_2)}$$
$$\frac{\zeta_1}{\Lambda_{10}} \bar{n}_2 = BN_2 - B\bar{n}_2 - N_2(\bar{n}_1+\bar{n}_2) + \overline{n_2(n_1+n_2)}$$

For $B, \mu_1/\lambda_{10}, \zeta_1/\Lambda_{10} \gg N_1, N_2$ we obtain

$$\bar{n}_1 = \frac{N_1}{1+\frac{\mu_1}{\lambda_{10}B}} \qquad \bar{n}_2 = \frac{N_2}{1+\frac{\zeta_1}{\Lambda_{10}B}}$$

This yields the mean free number of TFs and competitor C

$$n_{1,f} = \frac{N_1}{1+\frac{\lambda_{10}B}{\mu_1}} \qquad n_{2,f} = \frac{N_2}{1+\frac{\Lambda_{10}B}{\zeta_1}}$$

## 2. General formula for the target site search time and the occupation frequency of specific target sites

We use a multistate model in order to model specific binding sites. On these sites two effects can occur: unspecific binding and specific binding to the target site. In addition, the specific site can be empty with respect to the TF. We introduce three index sets

$$E = \{e_1, e_2, \ldots\} \quad D = \{d_1, d_2, \ldots\} \quad G = \{g_1, g_2, \ldots\} \qquad 15$$

E contains states in which no TF is bound, D contains states in which TFs are bound unspecifically, but not to the specific target site and G contains states in which the specific target site is occupied. The transition matrix T contains the transition rate constants between the states. The time evolution of the system is described by the Kolmogorov equation for the binding probabilities $p$

$$\dot{\vec{p}} = \mathbf{T}\vec{p} \qquad 16$$

The sum over all probabilities is normalized to 1.

We now calculate the target site search time. We define the search time as the time until any TF of the crowd binds to the target site, starting from a situation in which the specific site is unoccupied. Since we are interested in the first binding event we set all outgoing transitions of G to zero and thereby obtain the matrix $\mathbf{U}$. Next, we consider the flux of probabilities leaving the sets E,D

$$f = -\frac{d}{dt} \sum_{i \in E \cup D} p_i \qquad 17$$

Where $P_i$ denotes the probability to find the specific site in state i. In order to obtain this flux we need to solve the Cauchy problem of the Kolmogorov equations

$$\dot{\vec{p}} = \mathbf{U}\vec{p} \quad p_i = 0 \ \forall i \in D \cup G \quad p_i = p_{0,i} \ \forall i \in E \qquad 18$$

These initial conditions represent an empty specific site and therefore cover the whole target site search which consists of 3D diffusion and transition to the target site from the unspecifically bound state. The flux is the distribution of search times (Van Kampen, 1992). The expectation value of this distribution is obtained by

$$\tau = \int_0^\infty f(t) t \, dt = \int_0^\infty -t \frac{d}{dt} \sum_{i \in E \cup D} p_i \, dt \qquad 19$$

By partial integration we obtain

$$\tau = \left[ t \sum_{E \cup D} p_i \right]_0^\infty + \int_0^\infty \sum_{i \in E \cup D} p_i \, dt = \sum_{i \in E \cup D} Q_i \qquad 20$$

where

$$Q_i = \int_0^\infty p_i \, dt \qquad 21$$

Thus, the mean search time can be calculated by the integral over time of the probabilities $p_i$. These can be obtained by integration over both sides of Eq. 18 which yields

$$\vec{p}(\infty) - \vec{p}(0) = \mathbf{U}\vec{Q} \qquad 22$$

We now calculate the probability $p$ of the specific target site to be occupied (occupation frequency) by solving the steady state Kolmogorov equation

$$0 = \mathbf{T}\vec{p} \qquad 23$$

for $p$. We obtain the occupation frequency by summing over all target states G.

## Promoter activated by a single TF

We model the promoter (i.e. the specific site) with three states (Figure 5). The rate constants $\lambda$, $\lambda_1$ and $\lambda_2$ correspond to the transitions form state 1 to 3, 1 to 2 and 2 to 3 respectively. The rate constants denoted by $\mu$ represent the corresponding reverse reactions. The transition matrix of this three-state model is

$$\mathbf{T} = \begin{pmatrix} -\lambda - \lambda_1 & \mu_1 & \mu \\ \lambda_1 & -\mu_1 - \lambda_2 & \mu_2 \\ \lambda & \lambda_2 & -\mu - \mu_2 \end{pmatrix} \qquad 24$$

In order to represent the target site search problem, the states are divided into the sets

$$E = \{1\} \quad D = \{2\} \quad G = \{3\} \qquad 25$$

By solving Eq. 23 for $p_3$ we obtain the occupation frequency

$$p = \left[1 + \frac{m}{l}\left(1 + \frac{l_1}{m_1}\right)\right]^{-1} \qquad 26$$

By setting transitions that leave the specific target sites to zero we obtain

$$\mathbf{U} = \begin{pmatrix} -\lambda - \lambda_1 & \mu_1 & 0 \\ \lambda_1 & -\mu_1 - \lambda_2 & 0 \\ \lambda & \lambda_2 & 0 \end{pmatrix} \qquad 27$$

By using Eq. 22 and the initial conditions

$$\vec{p}(\infty) = (0,0,1) \quad \vec{p}(0) = (1,0,0) \qquad 28$$

we obtain the target site search time

$$\tau = Q_3 = \frac{\lambda_1 + \lambda_2 + \mu_1}{\alpha' \lambda} \quad \text{where} \quad \alpha' = \lambda_2 + \mu_1 + \frac{\lambda_1 \lambda_2}{\lambda} \qquad 29$$

*Promoter activated by two coactivating TFs*

We assume that both TFs independently bind to two specific target sites within the promoter (i.e. the specific site) with similar rate constants (Figure 2). We further assume that unspecific and specific sites within the promoter can be occupied by two cofactors at the same time, while the unspecific sites of the pool can only be occupied by one TF at a time. In order to represent the target site search problem, the states are divided into the sets

$$E = \{1\} \quad D = \{2,3,4,5\} \quad G = \{6\} \qquad 30$$

The transition matrix of the model for two coactivating TFs (see Figure 2) is

$$\mathbf{U} = \begin{pmatrix} -d_1 & \mu_1 & \mu & 0 & 0 & 0 \\ \lambda_1 & -d_2 & \mu_2 & 2\mu_1 & \mu & 0 \\ \lambda & \lambda_2 & -d_3 & 0 & \mu_1 & 0 \\ 0 & \lambda_1 & 0 & -d_4 & \mu_2 & 0 \\ 0 & \lambda & \lambda_1 & 2\lambda_2 & -d_5 & 0 \\ 0 & 0 & \lambda & 0 & \lambda_2 & 0 \end{pmatrix}$$

$$d_1 = \lambda_1 + \lambda$$
$$d_2 = \lambda_1 + \lambda_2 + \lambda + \mu_1$$
$$d_3 = \lambda_1 + \lambda + \mu + \mu_2$$
$$d_4 = 2\lambda_2 + 2\mu_1$$
$$d_5 = \lambda_2 + \mu_1 + \mu_2 + \mu$$

31

We calculated the search time by numerically solving the system of equations Eq. 31.

The probability *p* for the target site to be occupied (occupation frequency) is calculated according to Eq. 23 by detailed balance

$$p = p_6 = \frac{K^2}{1 + (1 + K + K_1)^2} \qquad 32$$

with equilibrium constants from Figure 2.

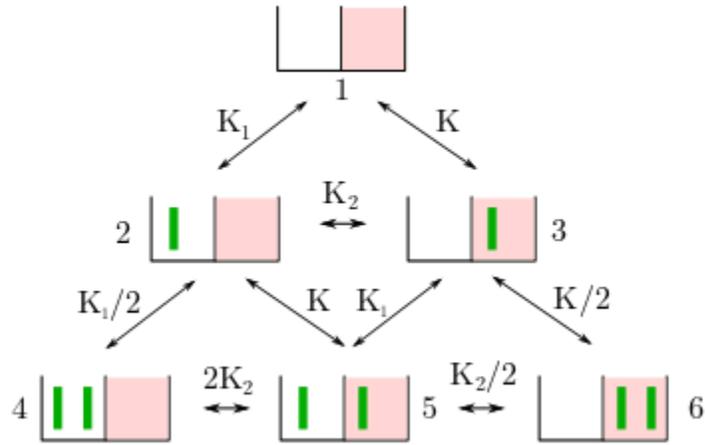

Figure 2: Six-state model of a specific site (e.g. a promoter) that is active if two cofactors are bound to the specific target site (red boxes). The unspecific site (white boxes) and the specific target site can contain two monomers at the same time. The equilibrium constants are written in terms of the ratios *bind/unbind* with the rates $K_l=\lambda_l/\mu_l$ $l=1,2$. The direct arrival at the target site with 3D diffusion of the protein is $K=\lambda/\mu$

## *Promoter activated by a single TF subject to inhibitory competition*

The competitor is assumed to inhibit binding of the TF within the promoter to unspecific sites or the specific target site (Figure 3). Unspecific sites of the pool can also be bound by either the competitor or the TF at a time. Since the unspecific and target sites can be empty or contain either a competitor or a TF we obtain nine states. In order to represent the target site search problem, these states are divided into the sets

$$E = \{1\} \quad D = \{2,3,4,6,7\} \quad G = \{5,8,9\} \qquad 33$$

Since the competitor presents a new kind of molecule with different association and dissociation rate constants, we introduce $\zeta_1$ and $\zeta$ for the unspecific and specific dissociation rate constants. Similar to the TF the competitor has arrival rates to unspecific and specific binding sites as well as a transition rate from unspecific binding on the specific site to specific binding on the target site. We denote these rates by $\Lambda_{10}$, $\Lambda_0$ and $\Lambda_2$ respectively. The transition matrix of the model for the competitor (see Figure 3) is

$$\mathbf{U} = \begin{pmatrix} -d_1 & \zeta & \zeta_1 & \mu_1 & 0 & 0 \\ \Lambda & -d_2 & \Lambda_2 & 0 & \zeta_1 & \mu_1 \\ \Lambda_1 & \zeta_2 & -d_3 & 0 & \zeta & 0 \\ \lambda_1 & 0 & 0 & -d_4 & 0 & \zeta \\ 0 & \Lambda_1 & \Lambda & 0 & -d_6 & 0 \\ 0 & \lambda_1 & 0 & \Lambda & 0 & -d_7 \end{pmatrix} \qquad 34$$

$$d_1 = \lambda_1 + \lambda + \Lambda_1 + \Lambda$$
$$d_2 = \lambda_1 + \Lambda_1 + \zeta_2 + \zeta$$
$$d_3 = \zeta_1 + \Lambda_2 + \Lambda + \lambda$$
$$d_4 = \mu_1 + \lambda + \Lambda + \lambda_2$$
$$d_6 = \zeta_1 + \zeta$$
$$d_7 = \mu_1 + \zeta$$

For simplicity we did not include the transitions to the set of states $G$ in Eq. 34. The search time is calculated by numerical matrix inversion of $\mathbf{U}$.

The probability $p$ for the target site to be occupied (occupation frequency) is calculated according to Eq. 23 by detailed balance

$$p = p_5 + p_8 + p_9 = \frac{1}{1 + K^{-1}(1+H)} \qquad 35$$

with equilibrium constants from Figure 3.

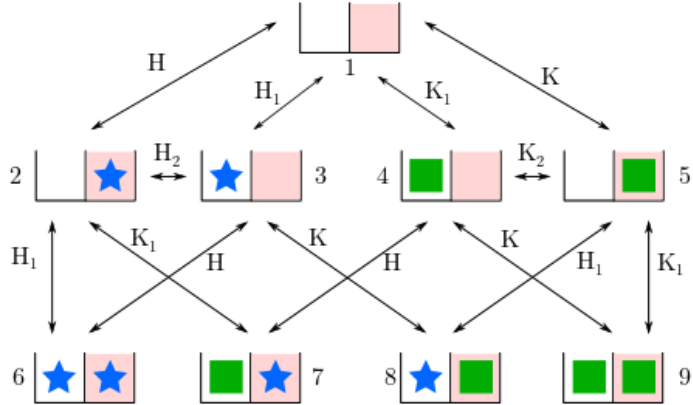

Figure 3: Nine state model of the specific site (e.g. a promoter) for binding of a TF (green square) and a competitor (blue star). On the promoter, the TF and the competitor can bind to the unspecific site (white boxes) and the specific target site (red boxes). Both species occupy these sites exclusively. Since each site can have three states, empty, competitor or TF bound, there are $3^2$ = 9 states of the promoter. The equilibrium constants are written in terms of the ratios *bind/unbind* with the rates $K_l=\lambda_l/\mu_l$; $H_l=\Lambda_l/\zeta_l$ $l=1,2$. The equilibrium constants $H2,K2$ are chosen such that the law of detailed balance is fulfilled. The direct arrival at the specific target site with 3D diffusion of the protein is $K=\lambda/\mu$ for the TF and $H=\Lambda/\zeta$ for the competitor.

# 3. Model for target site search by explicit 1D diffusion

We here model the TF target site search by explicitly accounting for 1D diffusion events on unspecific sites, interrupted by 3D diffusion until the specific target site is found (Bauer and Metzler, 2012; de la Rosa et al., 2010; Gerland et al., 2002; Halford and Marko, 2004; Klenin et al., 2006; Koslover et al., 2017; Krepel and Levy, 2016; Li et al., 2009; Mirny et al., 2009; Shvets and Kolomeisky, 2016; Slutsky and Mirny, 2004; Vonhippel and Berg, 1989) (Figure 4). The association rate constant to any base pair is $\lambda_0$. We model the sliding of the TF by single base pair sliding steps with transition rate constant $\alpha$ and an average sliding time of $\mu_1^{-1}$ (23). The transcription factor has one consensus DNA sequence to which it can bind specifically. As in (Li et al., 2009; Shvets and Kolomeisky, 2016), we include static roadblock molecules (crowders) in our model.

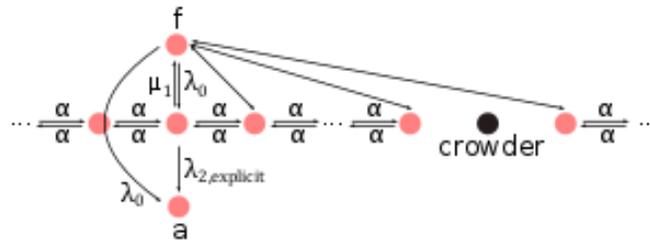

Figure 4: Model for TF target site search explicitly accounting for 1D and 3D diffusion based on a previous model (23). State f corresponds to the free TF. State a corresponds to the TF bound to its specific target site. The TF may arrive at any base pair of the genome by 3D diffusion and slides along unspecific DNA with sliding rate constant $\alpha$. If bound unspecifically to the target site the TF may switch to the specific binding mode with transition rate constant $\lambda_{2,expl}$. The TF may also arrive in the specifically bound conformation directly after 3D diffusion with rate constant $\lambda_0$. We accounted for crowding by periodically blocking sliding transitions between two sites every $M$ base pairs. With $R$ crowders the total scanned genome length thus is L=R·M.

We calculated the target site search time by applying our formalism of Materials and Methods, Section 2, to the initial value problem

$$E = \{f\} \quad D = \{-L/2,...,L/2\} \quad G = \{a\} \quad\quad 36$$

Where f denotes the unbound state, a denotes the occupied specific target site and L denotes the length of the scanned genome.

This leads to the result

$$\tau = \left[\left(\frac{1}{\lambda_0} + \frac{L-m}{\mu_1}\right)\cdot\left(\frac{\mu_1}{\lambda_{2,\exp l.}} + \frac{1}{m}\right) + \frac{m}{\lambda_{2,\exp l.}}\right]\cdot\left[1 + \frac{\mu_1}{\lambda_{2,\exp l.}} + \frac{1}{m}\right]^{-1}$$

$$m^{-1} = \frac{1+q^M}{1+q}\cdot\frac{1-q}{1-q^M} \qquad 37$$

$$q = 1 + \frac{\mu_1}{2\alpha} - \sqrt{\left(1+\frac{\mu_1}{2\alpha}\right)^2 - 1}$$

where M is the number of base pairs of one unspecific binding site. In order to find the sliding rate constant $\alpha$ between base pairs, we consider continuous diffusion, where $\mu_1 \to 0$, and omit crowding. In this case the TF can slide along DNA without restrictions. We obtain for the target site search time

$$\tau = \left(\frac{1}{\lambda_0} + \frac{L}{\mu_1}\right)\cdot\frac{1}{2\sqrt{\alpha/\mu_1}} \qquad 38$$

We compared this result to the formula

$$\tau = \left(\frac{1}{\lambda_0 L} + \frac{1}{\mu_1}\right)\cdot\frac{L}{n_D} \quad \text{where} \quad n_D = \frac{2}{l_{bp}}\sqrt{\frac{D}{\mu_1}} \qquad 39$$

derived in (Mirny et al., 2009) which includes the 1D diffusion constant $D$ measured experimentally in (Elf et al., 2007; Gorman et al., 2010; Kim and Larson, 2007) and the length of one base pair $l_{bp}$. The number $n_D$ is the number of visited base pairs per sliding event. Both Eqs. 38 and 39 yield the same results if we set

$$n_D = 2\sqrt{\frac{\alpha}{\mu_1}} = \frac{2}{l_{bp}}\sqrt{\frac{D}{\mu_1}} \qquad 40$$

This result enables us to obtain the sliding rate constant $\alpha$

$$\alpha = \frac{D}{l_{bp}^2} \qquad 41$$

Using $D$=0.01 µm²s⁻² (Gorman et al., 2010) and $l_{bp}$=0.34 nm we obtain $\alpha$=8.7·10⁴ s⁻¹. The optimally fast target site search time derived in (Mirny et al., 2009) requires a proportion of 1D diffusive search of 50%. Our model is consistent with this value in the limit case discussed above.

In Figure 6, our explicit diffusion model (Eq. 37) is plotted using the values N=100, B=10⁴ and $\lambda_0$=10⁻⁵ s⁻¹ as motivated in the main text. The rate for transition to binding on the specific target site, $\lambda_{2,expl}$=160 s⁻¹, was chosen such that the curves of the Diffusion model derived here agree with our new coarse-grained model of Eq. 47.

## 4. Application of the three-state model to eukaryotic TFs

According to our three-state model (Figure 5), a TF can dissociate from the specific target site using two distinct pathways: direct dissociation from the specific target site with $\mu$ or transition to the intermediate, unspecifically bound state with $\mu_2$ followed by dissociation with $\mu_1$. Therefore, the experimentally accessible specific dissociation rate constant represents an apparent rate constant determined by the eigenvalues $k$ of the Kolmogorov equations that include all dissociation pathways from the specific target site. In particular, the experimentally obtained specific dissociation rate constant does not directly correspond to the rate constant $\mu$. In order to apply our formalism to eukaryotic TFs, we therefore first need to calculate the specific dissociation rate constant $\mu$ from the experimentally measured apparent rate constant.

In order to calculate the eigenvalues $k$, we applied our formalism of Materials and Methods, Section 2, to the initial value problem where states 2 and 3 are occupied and state 1 is empty.

We find the equation

$$k^2 - (\lambda_2 + \mu_1 + \mu_2 + \mu)k + \mu_1\mu_2 + \mu(\mu_1 + \lambda_2) = 0 \qquad 42$$

for the eigenvalues.

Next, we obtain the specific dissociation rate constant $\mu$ using the measured rate constant $k$ by eliminating $\mu_2$ in Eq. 42 with the law of detailed balance and find

$$\mu = k \cdot \left(1 + M \frac{\lambda_2}{\mu_1} \frac{\mu_1 - k}{\lambda_2 + \mu_1 - k}\right)^{-1} \qquad 43$$

## Results

### Target site search time and occupation frequency of a gene

Besides binding to their specific target site, TFs have an, albeit lower, affinity to any sequence on DNA which gives rise to unspecific associations (Mustonen and Lassig, 2005; Slutsky et al., 2004; van Hijum et al., 2009). Since dissociation rate constants from DNA can readily be measured in living cells, we set out to develop a formalism for calculating the target site search time and occupation frequency of a gene based on dissociation rate constants. We thus considered a specific site (e.g. a promoter) embedded in a pool of unspecific binding sites (Figure 1) and translated this setting into a system of states where the promoter is decoupled from the pool of unspecific sites (Figure 5).

We first calculated the mean number of free TFs considering the pool of unspecific binding sites at equilibrium. Second, we solved the Kolmogorov equations for the specific site to obtain the time after which the first TF of the crowd occupied the specific target site (target site search time) and the probability of the specific target site to be occupied (occupation frequency). We considered three different promoter structures, a simple promoter bound by a single activating TF, a promoter activated by two independently binding TFs with similar kinetic rate constants and a promoter to which an additional inhibitory binding competitor can bind.

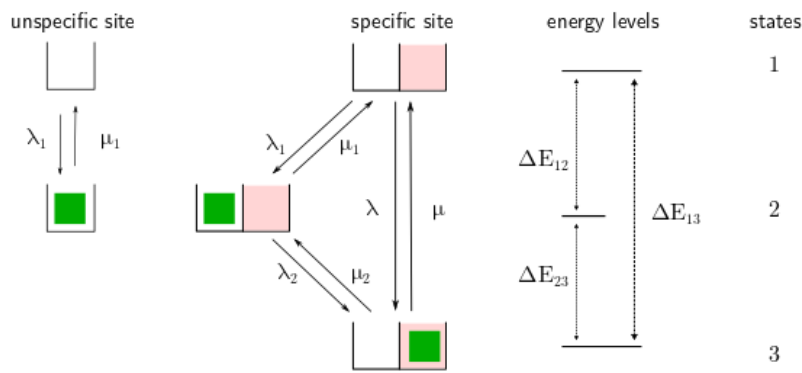

Figure 5: Three state model of TF-DNA binding. We consider a specific site for a TF (e.g. a promoter) including a specific target site (red box) and an adjacent unspecific site (white box), as well as a pool of uncoupled unspecific sites (left white box). The first state denotes free TFs, the second state denotes unspecifically bound TFs and the third state denotes TFs bound to the specific target site. We ascribe each state an energy level, energy differences are connected to the respective ratio of on- and off-rate constants by the law of detailed balance.

*Promoter activated by a single TF*

The simple promoter comprises the specific target site of the TF and surrounding unspecific DNA. Translated into our state-based formalism, the TF on the specific site (i.e. the promoter) has three states, free, unspecifically bound and bound to the specific target site (Figure 5). The promoter is embedded in a background of $B$ unspecific binding sites from which no direct transition to the promoter can occur. In our model, the TF can bind to its specific target site on the promoter either directly via 3D diffusion with association rate constant $\lambda$, or by first binding to the surrounding unspecific DNA with association rate constant $\lambda_1$ and a subsequent transition with rate constant $\lambda_2$. $\lambda_2$ is a cumulative rate including a potential conformational change associated with binding to the target site (Slutsky and Mirny, 2004) and a finite probability to find the target site when sliding on nearby unspecific sequences (Hammar et al., 2012). For the size of unspecific sites we chose a fixed number of 40 bp, corresponding to experimental findings (Gorman et al., 2010; Hammar et al., 2012). In vivo, the size of unspecific sites is limited by the presence of other DNA-bound molecules, not by the time a TF spends sliding on accessible DNA (Li et al., 2009; Shvets and Kolomeisky, 2016), justifying a fixed size independent of $\mu_1$.

The association rate constants $\lambda_1$ and $\lambda$ to any unspecific site or the specific target site depend on the number of free TFs, $n_f$ and the arrival rates of a single TF $\lambda_{10}$ and $\lambda_0$ on unspecific and specific sites. This number in turn depends on the dissociation rate constant of unspecific sites, $\mu_1$, according to

$$\lambda_1(\mu_1) = n_f(\mu_1) \lambda_{10} \qquad \lambda(\mu_1) = n_f(\mu_1) \lambda_0 \qquad \qquad 44$$

Here, we neglect a dependence on the dissociation rate constant of the specific target site since only one such site exists, which is not occupied during target site search. $\lambda_0$ and $\lambda_{10}$ denote the diffusion limited arrival rate constants.

Association rate constants and dissociation rate constants are coupled by the law of detailed balance

$$\frac{\lambda_{10}}{\mu_1} \times \frac{\lambda_2}{\mu_2} = \frac{\lambda_0}{\mu} \qquad \qquad 45$$

We first calculated the number of free TFs $n_f$ (Material and Methods, Section 1). If the total number of TFs $N$ is much smaller than $B$ and the ratio $\mu_1/\lambda_{10}$, we find for $n_f$

$$n_f(\mu_1) = \frac{N}{1 + \frac{\lambda_{10} B}{\mu_1}} \qquad \qquad 46$$

This equation shows that the number of free TFs significantly decreases if the number of unspecific sites is large and unspecific binding is favoured by a large ratio $\lambda_{10}/\mu_1$. For the following discussion we used the exact result for $n_f$ (Eq. 4 derived in Materials and Methods, Section 1) and evaluated it numerically.

Next, we calculated the target site search time $\tau$ of the gene. We assumed an equilibrium distribution of free and unspecifically bound TFs in the pool of unspecific binding sites, given an unoccupied promoter, and used the Kolmogorov equations to calculate the time until the first TF bound the target site (Materials and Methods, Section 2). Using the number of free TFs, we found for the target site search time (Figure 6)

$$\tau = \frac{\lambda_1 + \lambda_2 + \mu_1}{\alpha' \lambda} \quad \text{where} \quad \alpha' = \lambda_2 + \mu_1 + \frac{\lambda_1 \lambda_2}{\lambda} \qquad \qquad 47$$

where $\lambda_2$ is the transition from unspecific binding on the specific site to binding to the specific target site. The search time exhibits a minimum corresponding to an optimally fast target search process, as for small $\mu_1$ TFs are drawn towards unspecific DNA and the free pool of searching TFs becomes minimal while for large $\mu_1$ the benefit of 1D diffusional scanning of DNA is negligible. This result is consistent with previous studies of the TF target site search problem that explicitly included 1D diffusion (Berg et al., 1981; Halford and Marko, 2004; Mirny et al., 2009; Winter et al., 1981).

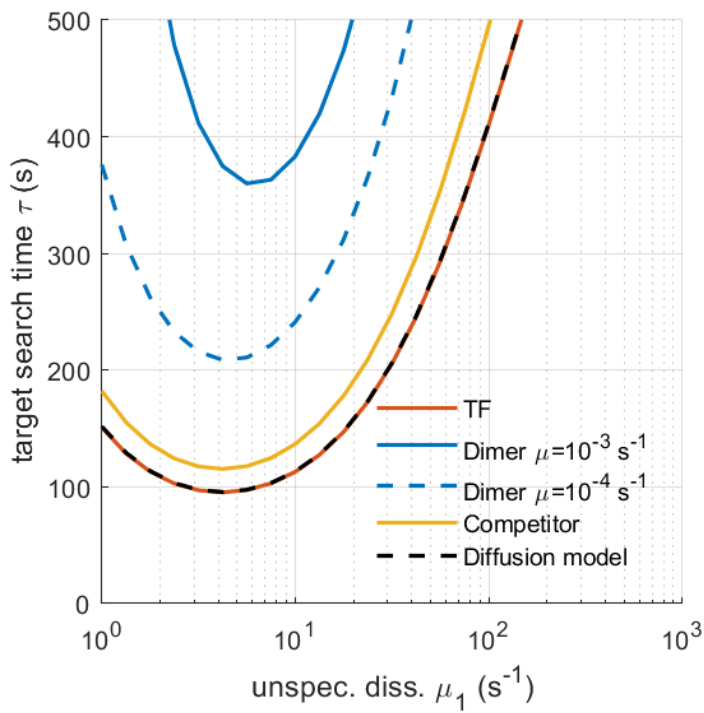

Figure 6: : Comparison of the target site search time $\tau$ of the gene as function of the unspecific dissociation rate constant $\mu_1$ between a promoter activated by a single TF (TF, red line), a promoter activated by a dimer (Dimer $\mu=10^{-3}$ s$^{-1}$, solid blue line), activated by a dimer with lower specific dissociation rate constant (Dimer $\mu=10^{-4}$ s$^{-1}$, dashed blue line) and when an inhibitory competitor of the TF is present (Competitor, yellow line). In addition, the TF target site search time calculated from a model explicitly accounting for 1D diffusion is indicated (Diffusion model, dashed black line). Simulation parameters are: N=100, C =100, B=10$^4$, $\lambda_{10}=\Lambda_{10}=4\cdot10^{-4}$ s$^{-1}$, $\lambda_0=\Lambda_0=10^{-5}$ s$^{-1}$, $\lambda_2=\Lambda_2=4$ s$^{-1}$, $\zeta_1=1$ s$^{-1}$, $\zeta=10^{-3}$ s$^{-1}$.

To validate our simplified calculation of the target site search time, we compared our result with the one from a microscopic model explicitly treating 1D diffusion in presence of static roadblocks that limit the sliding distance of TFs (Li et al., 2009; Shvets and Kolomeisky, 2016) (Materials and Methods Section 3, Figure 4). Both models agree very well (Figure 6). Since our simplified state based approach allows application to experimentally measured TF rate constants and enables straightforward incorporation of more complex promoter structures, we used it hereafter.

We calculated the probability $p$ of the specific target site to be occupied (occupation frequency) of the gene by eliminating $\lambda_2/\mu_2$ with Eq. 45 and solved the Kolmogorov equations (Materials and Methods, Section 2). We found for the occupation frequency (Figure 7)

$$p = \left[1 + \frac{m}{l}\left(1 + \frac{l_1}{m_1}\right)\right]^{-1} \qquad 48$$

This parameter increases monotonically with $\mu_1$ and asymptotically approaches a maximum determined by the equilibrium constant of specific binding, since $\mu_1$ directly determines the number of free TFs and thus the association rate constant to specific target sites (Eq. 44).

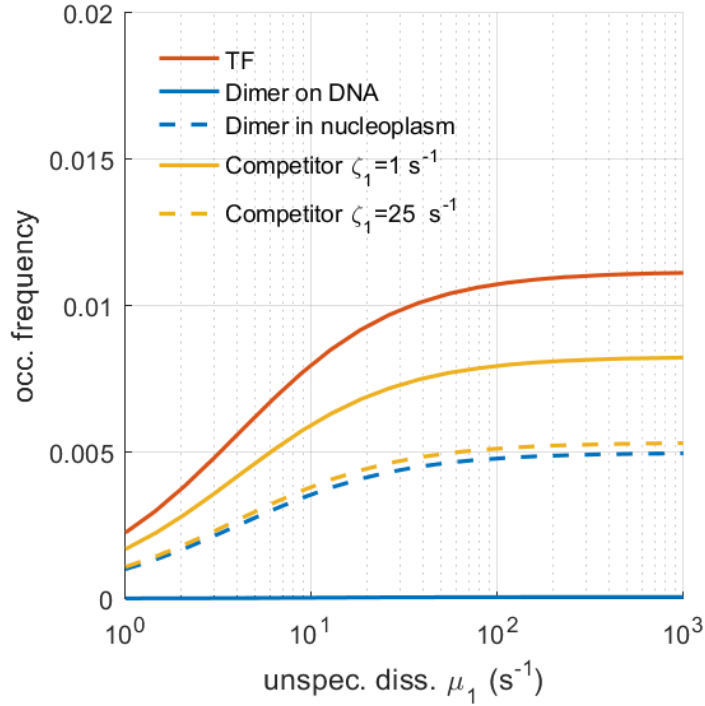

Figure 7: Comparison of occupation frequency $p$ as function of the unspecific dissociation rate constant $\mu_1$ between a promoter activated by a single TF (TF, red line), a promoter activated by a dimer formed on DNA (Dimer on DNA, blue line) and by a dimer formed in the nucleoplasm (Dimer in nucleoplasm, dashed blue line), when an inhibitory competitor of the TF is present (Competitor $\zeta_1=1$ s$^{-1}$, yellow line) and when the competitor has a larger unspecific dissociation rate constant (Competitor $\zeta_1=25$ s$^{-1}$, dashed yellow line). Simulation parameters are: N=100, C=100, B=10$^4$, $\lambda_{10}=\Lambda_{10}=4\cdot 10^{-4}$ s$^{-1}$, $\lambda_0=\Lambda_0=10^{-5}$ s$^{-1}$, $\lambda_2=\Lambda_2=4$ s$^{-1}$, $\mu=10^{-1}$ s$^{-1}$, $\zeta=10^{-3}$ s$^{-1}$.

When comparing Figures 6 and 7, it becomes obvious that the unspecific dissociation rate constant $\mu_1$ at which the target site search time is minimal does not necessarily correspond to a high occupation frequency, indicating that target site search time and occupation frequency might not simultaneously be optimized for a given TF. However, since the occupation frequency depends on the specific dissociation rate constant while the target site search time does not, both parameters might be optimized by adjusting the specific dissociation rate constant.

*Promoter activated by two coactivating TFs*

To model gene activation by two independently binding TFs of similar type, the promoter comprises two specific target sites, one for each TF, surrounded by unspecific DNA. Both TFs are assumed to bind independently with similar rate constants but activate the gene only if bound specifically at the same time as homo- or heterodimers. Translated into our state-based formalism, the TFs on the specific site have six states (Materials and Methods Section 2, Figure 2). We deliberately omitted binding interactions between both TFs (Geisel and Gerland, 2011), to focus on the contributions of DNA binding to target site search and occupation frequency and preserve the comparison to a single TF.

To calculate the target site search time of the promoter activated by two TFs, we solved the Kolmogorov equations of the dimeric system numerically (Figure 6 and Materials and Methods, Section 2). For the occupation frequency, we compared dimerization exclusively on DNA with dimerization already in solution (Figure 7).

While for a single TF the target site search time is independent of the specific dissociation rate constant, two independently binding coactivating TFs simultaneously find the specific target site the faster, the smaller the specific dissociation rate constant is. This reflects the fact that both factors have to wait for each other on the specific target site. Accordingly, the probability of simultaneous occupation of the target site is very low for two independently binding TFs. This occupation frequency increases if dimerization is allowed to occur already in the nucleoplasm.

*Promoter activated by a single TF subject to inhibitory competition*

To model gene activation in the presence of an inhibitory binding competitor, we introduced a second species that is able to bind to the same unspecific and specific binding sites as the TF, but with own dissociation rate constants. Once occupied by the competitor, a binding site is sterically blocked for binding of the TF. Translated into our state-based formalism, the system has nine states (Materials and Methods Section 2, Figure 3).

We again numerically solved the Kolmogorov equations of the system with binding competitor to obtain the target site search time of the gene. The target site search time of the TF increases significantly if the unspecific dissociation rate constant $\zeta_1$ of the competitor increases (Figure 8a). In this case more competitor is available to block the target site. Consequently, the occupation frequency of the gene drops under these conditions (Figure 8b). If the occupation of the specific target site by the competitor is modulated with the specific dissociation rate constant $\zeta$ of the competitor, a small value for $\zeta$, i.e. a high occupation, leads to a similar effect (Figure 8c and d). These findings agree well with (Zabet and Adryan, 2013), where an increased coverage of DNA by mobile competitors and with that of the specific binding site was achieved by increasing competitor abundance.

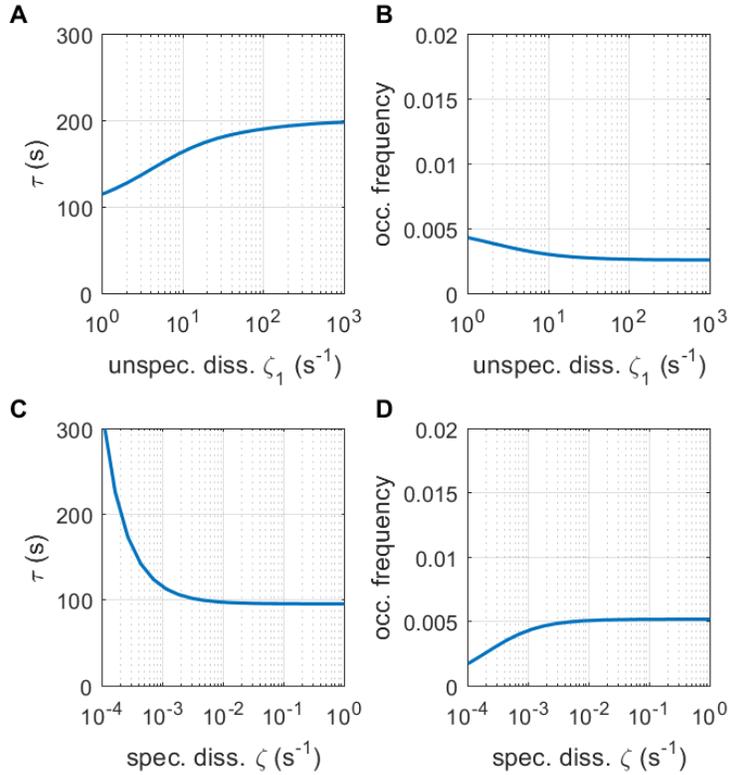

Figure 8: Influence of an inhibitory binding competitor. (A) Target site search time $\tau$ of the gene and (B) probability $p$ of the promoter to be occupied by a TF (occupation frequency) as function of the unspecific dissociation rate constant $\zeta_1$ of the competitor. (C) $\tau$ and (D) $p$ as function of the specific dissociation rate constant $\zeta$ of the competitor. Simulation parameters are: N=100, C =100, B=10$^4$, $\lambda_{10}=\Lambda_{10}=4\cdot10^{-4}$ s$^{-1}$, $\lambda_0=\Lambda_0=10^{-5}$ s$^{-1}$, $\lambda_2=\Lambda_2=4$ s$^{-1}$, µ=10$^{-1}$ s$^{-1}$, $\zeta$=10$^{-3}$ s$^{-1}$.

### TFs are optimized for fast target site search but not for high occupation frequency

Since both target site search time and occupation frequency depend on the unspecific dissociation rate constant $\mu_1$, but only the occupation frequency depends on the specific dissociation rate constant $\mu$, we tested under which conditions target site search and occupation frequency assumed their respective optima, fast target site search and high occupation frequency (Figure 9). To compare both quantities, we chose parameter values according to the physiological conditions in an eukaryotic cell. We varied the unspecific dissociation rate constant $\mu_1$ in the interval 10$^0$ s$^{-1}$ to 10$^3$ s$^{-1}$ (Ball et al., 2016; Caccianini et al., 2015; Elf et al., 2007; Morisaki et al., 2014) and the specific dissociation rate constant $\mu$ from 10$^0$ s$^{-1}$ to 10$^{-4}$ s$^{-1}$ (Agarwal et al., 2017; Caccianini et al., 2015; Chen et al., 2014; Gebhardt et al., 2013; Hammar et al., 2014; Mazza et al., 2012; Speil et al., 2011; Sugo et al., 2015). The arrival rate constant is given by the 3D diffusion limit under physiological salt concentrations, $\lambda_0$=10$^6$ (Ms)$^{-1}$ (Mirny et al., 2009), which results in $\lambda_0$=10$^{-5}$ s$^{-1}$ in a typical cell volume of 100µm$^3$. The arrival rate constant to the unspecific site scales with the size of unspecific sites of 40bp to $\lambda_{10}$=4·10$^{-4}$ s$^{-1}$. We adjusted the transition rate constant $\lambda_2$=4 s$^{-1}$ in accordance with a low probability to bind the specific target site when scanning nearby unspecific DNA (Hammar et al., 2012). In this case the rate limiting step of specific DNA-binding is the 3D diffusion to the specific site. Finally, we assumed that the promoter is embedded in a background of $4\cdot10^5$ bp of unspecific DNA, resulting in B=$10^4$ unspecific binding sites. This value revealed a fraction of DNA-associated TFs of 20%-40% (Materials and Methods Section 1, Eq. 6) reported by experiments (Chen et al., 2014; Gebhardt et al., 2013; Mazza et al., 2012). This fraction corresponds to the fraction of time the TF spends in 1D search. We found that a common optimum of target site search time and occupation frequency is only achieved for large unspecific and small specific dissociation rate constants (Figure 9).

We next applied our model to various eukaryotic TFs. We calculated their occupation frequencies using published values of unspecific and specific dissociation rate constants measured in vivo, and plotted these onto the theoretically derived occupation frequency (Figure 9 and Table 1). It is important to note that experimentally accessible dissociation rate constants are apparent rate constants that first have to be converted to the specific dissociation rate constant of the three-state model (Materials and Methods, Section 4). To be able to compare the different TFs, we assumed that dimeric TFs such as GR already dimerized in solution, and considered a ratio of 100 TFs per specific target site, which is a good approximation for many eukaryotic TFs binding to hundreds of specific target sequences (Geiger et al., 2012). However, we note that calculation of the target site search time requires knowledge about the number of TFs present in the cell and the ratio of specific to unspecific binding sites, which will be specific to each TF. We found that all considered TFs are optimized for a fast target site search time. In contrast, the probability of most TFs to occupy their specific target site is low. Only TFs with very small specific dissociation rate constants such as CTCF have a high occupation frequency.

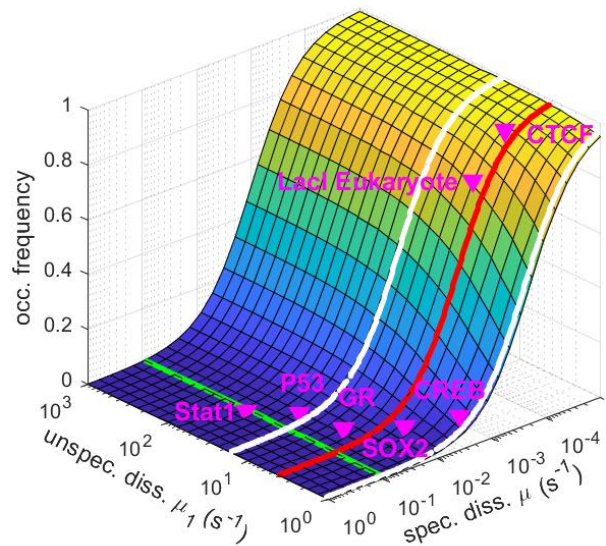

Figure 9: Occupation frequency of the promoter by a TF as function of the unspecific ($\mu_1$) and the specific ($\mu$) dissociation rate constants. The values $\mu_1$ of minimal (red line) and 1.5 times the minimal target site search time (white lines) are indicated. The green line at $\mu=10^{-1}$ s$^{-1}$ indicates the cross section of the occupation frequency displayed in Figure 7. Symbols correspond to occupation frequencies of various TFs calculated from unspecific and specific dissociation rate constants measured by fluorescence microscopy in live cells. Simulation parameters are: N=100, B=$10^4$, $\lambda_{10}$ =$4\cdot10^{-4}$ s$^{-1}$, $\lambda_0$ =$10^{-5}$ s$^{-1}$, $\lambda_2$= 4 s$^{-1}$.

| TF | $\mu_1$ (s$^{-1}$) | $k$ (s$^{-1}$) | $\mu$ (s$^{-1}$) | source |
|---|---|---|---|---|
| Sox2 | 1.3 | 0.08 | 0.003 | (Chen et al., 2014) |
| LacI (mammalian cell) | 5.6 | 0.003 | 0.0002 | (Caccianini et al., 2015; Hammar et al., 2014) |
| CREB | 2.7 | 0.2 | 0.009 | (Sugo et al., 2015) |
| P53 | 20 | 0.3 | 0.04 | (Mazza et al., 2012) |
| Stat1 | 50 | 0.4 | 0.1 | (Speil et al., 2011) |
| GR | 6.7 | 0.5 | 0.03 | (Gebhardt et al., 2013; Groeneweg et al., 2014) |
| CTCF | 5 | 0.001 | 0.00005 | (Agarwal et al., 2017) |

Table 1: Unspecific ($\mu_1$) and apparent specific ($k$) dissociation rate constants of transcription factors measured in live cells. The apparent dissociation rate constant $k$ is converted to the specific dissociation rate constant of the three-state model, $\mu$, using Eq. 43.

## Discussion

We have introduced a treatment of TF target site search that is based on TF-DNA dissociation rate constants accessible with live cell measurements and thus readily applicable to experiments. To achieve this, we replaced explicit 1D diffusion of the TF with its dissociation rate constant from unspecific sites. We have shown that this replacement is able to model the search process in presence of explicit diffusion and static roadblock molecules or crowding molecules (Koslover et al., 2017; Krepel and Levy, 2016; Li et al., 2009; Morelli et al., 2011; Shvets and Kolomeisky, 2016) to good approximation. Our formalism thus yields very transparent and intuitively accessible analytical solutions for the target site search problem.

When applying our model to experimentally measured dissociation rate constants of eukaryotic TFs, we find that TFs indeed operate at the optimum of a fast target site search, but high occupation frequencies are not necessarily achieved in cells. This has also been observed for Drosophila TFs (Zabet and Adryan, 2015). While the low occupation frequency could in principle be counteracted by a high specific dissociation rate constant, such a strategy does not seem to occur for most TFs. Another way to increase the occupation frequency is to increase the total number of TFs. This effect becomes particularly important if the TF number is comparable to the number of unspecific binding sites. A possible reason for suboptimal occupation frequency could arise from the stability paradox (Gerland et al., 2002; Slutsky and Mirny, 2004), which states that the energy gap between unspecific and specific binding cannot be arbitrarily high and a compromise between both values might need to be attained. On the other hand, these particular choices of specific dissociation rate constants may also be advantageous for the functioning of TFs. Most eukaryotic TFs act not only on one promoter but regulate hundreds of genes and contribute to different signalling cascades. In such a genetic network short specific binding times ensure that always a large fraction of molecules is available in solution and the time to find a specific target site is kept low. This enables fast reactions to changing signalling environments. In contrast, the long specific binding time measured for the chromatin architecture protein CTCF (Agarwal et al., 2017) leads to a high fraction of occupied target sites and a considerable reduction of free protein. Thus, the target site search time will be high once a CTCF molecule dissociated and CTCF-mediated chromatin loops will only reform slowly.

Optimizing for fast target site search and high occupation frequency may not be the only constraint in the evolution of dissociation rate constants. Given the complexity of gene regulation, target site search might not be the rate limiting step. Additionally, other requirements such as fast reaction to signals or a kinetic separation of monomer and dimer function as discussed below may shape TF dissociation rate constants.

In addition to being easily applicable to experimentally accessible data, our target search formalism simplifies the search problem by mathematically decoupling the discussion of TF binding-modalities at the specific site (e.g. different promoter structures) from unspecific binding to the pool of unspecific binding sites. Accordingly, our initial condition for the target site search is equilibrium between free and unspecifically bound TFs, and we calculated the time after which the first TF of the crowd binds to the specific target site. This is in contrast to the common approach considering the full search pathway of a single TF. However, since the transient process toward the initial equilibrium converges with a rate of B·$\lambda_1$, which is much faster than the target site search time, our assumption for the initial condition barely influences the search time. Thus, while being coarse-grained, our model is still general and powerful. It can be further extended to include the topology of DNA and crowding effects in the nucleoplasm (Hu et al., 2006; Kamar et al., 2017; Morelli et al., 2011), by appropriate changes to the arrival and dissociation rate constants. In particular, protein-protein interactions could easily be incorporated using additional states. Our model may in addition accelerate the simulation of a background of unspecific binding sites, by replacing modelling of myriads of explicit 1D sliding states by the unspecific dissociation rate. A similar technique was applied to scale and accelerate the simulation of the entire facilitated diffusion process (Zabet, 2012; Zabet and Adryan, 2012).

A state-based formalism was previously used in (Schoech and Zabet, 2014) to investigate the buffering effect of facilitated diffusion on gene expression noise. This state-based model combines 1D and 3D diffusion in a single association rate constant to the specific target site. Rebinding events to the target site by sliding are explicitly considered by an additional state and shown to be equivalent to an effective dissociation rate constant. In our approach, we used the state-based formalism to model the search time of the TF for its specific target site in more detail by including a state for unspecific binding.

The non-monotone dependency of the target site search time on the duration of 1D sliding, respectively the unspecific dissociation rate constant in our approach, has been discussed comprehensively in previous studies (Bauer and Metzler, 2012; de la Rosa et al., 2010; Gerland et al., 2002; Klenin et al., 2006; Mirny et al., 2009; Vonhippel and Berg, 1989). Here, we focused on how the search process adapts to more complex promoters that involve coactivators and inhibitory competitors. The microscopic diffusion processes constrained by position and overlap of binding sites for competitors and TFs were previously motivated and investigated (Ezer et al., 2014).

We considered the limit cases of exclusive dimerization in the nucleoplasm and exclusive dimerization on the target site in absence of a stabilizing protein-protein interaction (Geisel and Gerland, 2011). This allows direct comparison of search times and occupation frequencies of two coactivating TFs with a single TF. We found that two independently binding TFs prolong the target site search time significantly. Additionally, the unspecific dissociation rate constant $\mu_1$ for an optimal target site search time increases. This increase is accompanied by a decrease of the transition rate constant $\mu_2$ from a specifically to an unspecifically bound TF due to the law of detailed balance, meaning that the two TFs have to wait for each other on the specific target site to achieve fastest gene response times. Alternatively, dimeric TFs may wait for each other by increasing the specific dissociation rate constant $\mu$. Whether dimeric TFs generally exhibit larger dissociation rate constants from unspecific DNA or smaller dissociation rate constants from specific target sites compared to solely activating TFs is subject to future studies.

For inducible dimers capable to dimerize in solution, such as nuclear receptors (Horwitz et al., 1996), the non-linear increase of target site search time at low dissociation rate constants may lead to a kinetic separation of monomer and dimer function. While uninduced monomers in principle could simultaneously arrive at the promoter, potentially evoking background activation, such a search mechanism is much slower than the one for dimer molecules that effectively bind as one TF. This turns background activation of uninduced monomeric molecules inefficient. The occupation frequency of the promoter activated by two TFs drops significantly compared to the case of a single TF. The lower gene activation capability arises due to the more complex activation mechanism and could be accounted for by increasing the monomer concentration or by allowing dimerization not only on DNA but already in the nucleoplasm. This constitutes an additional advantage for solution-based dimerization and sheds new light on the question whether for example nuclear receptors bind to DNA as monomers or dimers (Ong et al., 2010).

We have discussed the influence of a competitor that inhibits binding to the specific target site. The modulatory effect of an inhibitory competitor on the target site search time and occupation frequency of a TF can be mainly ascribed to blocking of the specific site rather than preventing the TF from scanning unspecific DNA sequences common to all DNA binding proteins. Thus, an inhibitory competitor will indeed only alter the activity of a certain gene without changing the search times or occupation frequencies of other genes (Clauß et al., 2017). In contrast, an unspecific competitor will mainly affect TF occupation at unspecific sites, turning occupation of the target site more susceptible to TF abundance (Zabet et al., 2013).

## Acknowledgements
We thank Matthias Reisser for helpful discussions. This work was supported by the German Research Foundation [GE 2631/1-1 to J.C.M.G.] and the European Research Council (ERC) under the European Union's Horizon 2020 Research and Innovation Programme [637987 ChromArch to J.C.M.G.].

## Author Contributions
J.H. and J.C.M.G. designed the study; J.H. performed calculations; J.H. and J.C.M.G. wrote the manuscript.